\documentclass{aa}  
\usepackage{graphicx}
\usepackage{txfonts}
\usepackage{amsmath}
\usepackage{amssymb}
\usepackage{multicol}
\usepackage{natbib}
\usepackage{comment}
\usepackage[colorlinks=true,allcolors=blue]{hyperref}
\usepackage{hyperref}
\newcommand{\ergs}[0]{\rm erg\,\rm s^{-1}}

\newcommand{\eflux}[0]{{\rm erg}\,{\rm cm}^{-2}\,{\rm s}^{-1}}
\defcitealias{longo25}{L25}
\defcitealias{bosch23}{BR23}

\begin{document} 
\title{Broadband non-thermal emission predictions based\\ on 3D simulations of supernova explosions within active galactic nuclei jets}

    \titlerunning{Non-thermal emission}
   
   \author{B. Longo\inst{1}
          \and
          V. Bosch-Ramon
          \inst{2}
          \and
          M. Perucho
          \inst{1,3}
          \and
          J. M. Mart\'{\i}
          \inst{1,3}
          }

    \authorrunning{Longo, Bosch-Ramon, Perucho \& Mart\'{\i}}
          
   \institute{ Departament d'Astronomia i Astrof\'{\i}sica, Universitat de Val\`encia, C/ Dr. Moliner,
    50, 46100, Burjassot, Val\`encia, Spain.
             \and
            Departament de F\`isica Qu\`antica i Astrof\'isica, Institut de Ciències del Cosmos (ICC), Universitat de Barcelona (IEEC-UB), Mart\'i i Franqu\'es 1, E08028 Barcelona, Spain.
         \and
             Observatori Astron\`omic, Universitat de Val\`encia, C/ Catedr\`atic Jos\'e Beltr\'an 2, 46980, Paterna, Val\`encia, Spain.
             }
             
   \date{}
  \abstract 
   {Thermonuclear and core-collapse supernovae are statistically expected to happen within extragalactic jets, as these jets cross the inner regions of their host galaxies, and supernova ejecta--jet interactions can potentially lead to strong non-thermal activity.}
   {We model the non-thermal emission from a supernova interacting with a jet on kiloparsec scales, using detailed numerical simulations of the interaction.}
   {Using a post-processing method, we applied a non-thermal model of the acceleration and transport of electron and their synchrotron and inverse Compton emissions, simplified and tuned for flexibility and computational speed, to the results of a three-dimensional relativistic hydrodynamics simulation of the ejecta of a supernova interacting with an extragalactic jet.}
   {For typical magnetic field strengths, a jet power $10^{44}\,\ergs$, and an electron energy distribution characteristic of diffusive shock acceleration, we predict: synchrotron emission that peaks in X-rays but reaches up to 100~MeV, with an overall luminosity of a few times $10^{41}\,\ergs$; and emission from inverse Compton scattering mostly off galactic infrared photons, peaking at $\sim 1$~TeV and with a luminosity up to $\sim 10^{40}\,\ergs$. We also computed the evolution of the emission, and the distribution of the flux in the sky at its maximum, for five different energy bands: radio, optical, X-rays, high-energy and very high-energy $\gamma$-rays. The emission is roughly steady on a few 1000~yr timescales, with an extension on arcsec~scales for a source in the local universe.}
   {For intermediate jet powers, radio, optical and X-ray fluxes from supernova ejecta-jet interactions may be detectable for distances $\lesssim 100$~Mpc. The non-thermal luminosity budget and local conditions associated with these interactions could be enough to explain knots seen in kiloparsec scale jets at different energies. These conditions might also be suitable for the production of heavy ultra high-energy cosmic rays, but photon and neutrino generation from hadronic processes would be minor on kiloparsec scales. The leptonic $\gamma$-ray fluxes are modest but might non-negligibly contribute in TeV for the nearest or more powerful jet sources.
   }
   \maketitle


\section{Introduction}

Born from the accretion onto supermassive black holes (SMBH) at the centre of active galaxies, jets from active galactic nuclei (AGN) are collimated outflows that propagate from the innermost regions of the host galaxy, at sub-parsec scales, to the intergalactic medium, even up to megaparsec scales, where they terminate in forming lobes filled with hot plasma \citep[e.g.][]{blandford79,begelman84,blandford19,dabhade20,andernach25}. Extragalactic jets have been extensively observed in radio, optical, X- and $\gamma$-rays \citep[][]{bri84,harris02,perlman06,harris06,kov07,god12,hess20}, when they can be resolved enough to be distinguished from the host galaxy and the AGN itself. In certain cases, like the radio-galaxies Centaurus~A and M87, observations at different energies have unveiled the presence of knots located in the jet on 0.1--1~kpc scales \citep[e.g.][]{kraft01,wilson02,kraft02,hardcastle03,ly07,worral08,goodger10,rod25}. Acceleration of particles in shocks is commonly invoked to explain these structures \citep{kirk87,meisenheimer96,kirk97,meisenheimer03}. The inner galactic regions are known to contain large amounts of gas, stars and related objects (e.g. \citealt{burbidge70}; \citealt{bednarek97}; \citealt{kauffmann03}; \citealt{hub06}; \citealt{shao10}; \citealt{wykes15}). Stars, in particular, are expected to be present inside the jet either because the latter engulfed them when they first crossed the host galaxy, or because galactic motion brought them inside the jet, so interactions between outflows from these stars and the jet flow are unavoidable \citep[e.g.][]{komissarov94,bowman96,hub06,bosch12,perucho14,perucho17,castillo2021}. These interactions can potentially channel non-negligible amounts of jet energy into non-thermal processes \citep[e.g.][]{fed96,bednarek97,bed99,barkov10,bosch12,araudo13,bed15,delacita16,2025A&A...693A.270F}, while leaving detectable signs in the jet dynamical structure \citep[as suggested by, e.g.][]{mul14}.

Due to the presence of stars of all masses and evolutionary stages, supernova (SN) explosions can occur inside an extragalactic jet \citep{blkosnr,fed96,bed99}. These SN explosions would be most likely core-collapse in the case of galaxies with significant star formation \citep{vieyro19}, or thermonuclear in galaxies with little star-formation activity \citep{tor19}. In these events, the SN ejecta, while being impacted by the jet, creates a bow-shaped shock that can lead to significant non-thermal activity, as proposed by \citet{vieyro19} and \citet{tor19}, and further explored in \cite{bosch23} (\citetalias{bosch23} hereafter) and \citet{he26}. However, these non-thermal studies used simplified hydrodynamics models, and realistic 3D simulations were lacking. \citet{longo25} (\citetalias{longo25} hereafter) carried out a set of high-resolution 2D and 3D relativistic hydrodynamics (RHD) simulations of the interaction between an SN ejecta and a stationary extragalactic jet. These simulations showed that the fast ejecta expansion, strongly enhanced by heating due to the jet impact, leads to a strong shock that can cover most of the jet section for typical parameters. Such a wide shock can convert a significant fraction of the kinetic energy into non-thermal energy, as it is well known for strong shocks. In this work, profiting from results obtained by \citetalias{longo25}, we characterise the non-thermal particles accelerated in a SN~ejecta--jet interaction, and compute the subsequent synchrotron and Inverse Compton (IC) emission using detailed RHD information from the simulations. The paper is structured as follows: in Sect.~\ref{simu}, we summarise the 3D RHD simulations whose results were used in this work; in Sect.~\ref{model}, we detail the light-cost non-thermal model used in the calculations of non-thermal processes; in Sect.~\ref{resu}, we present the obtained results; and in Sect.~\ref{disc}, we discuss the implications of this work and summarise our main results.

\section{Simulations}\label{simu}

\citetalias{bosch23} estimated that $\sim 1$\% of all AGN jets may host a SN~ejecta. Thus, motivated by the fact that these events can happen with non-negligible rates, \citetalias{longo25} carried out high-resolution 3D RHD simulations in Cartesian coordinates of an SN explosion within an AGN jet on kiloparsec scales. The SN ejecta was initially modelled as an overpressured uniform plasma sphere, with a mass of $M_{\rm SN}=2$~M$_\odot$, temperature $T_{\rm SN}=10^9$~K, and energy $E_{\rm SN}=10^{51}$~erg, expanding against the flow of a jet with radius $R_{\rm j}=100$~pc, power $L_{\rm j}=10^{44}$~erg~s$^{-1}$, Lorentz factor $W_{\rm j}=2$, and temperature $T_{\rm j}=2\times 10^{11}$~K. The jet parameters adopted by \citetalias{longo25}, that is, a moderately relativistic and hot, but still highly supersonic, proton-electron plasma, were justified by the fact that for not very powerful AGN, kiloparsec-scale jets should have been significantly mass-loaded within the inner regions of the galaxy, which is expected based on observational and theoretical arguments \citep[see, e.g.,][and references therein]{perucho14,castillo2021}. We assumed a stationary jet, but this may not be the case on the timescales of interest (which is discussed in Sect.~\ref{disc}). The size of the computational grid in each Cartesian direction was taken as $L_x=L_y=L_z=100\,R_{\rm SN}$, where $-50\,R_{\rm SN}\le x,y,z\le 50\,R_{\rm SN}$, with $R_{\rm SN}=1.1$~pc being the initial ejecta radius
and equal to eight cells. All the parameter values of the simulation given above are listed in Table~\ref{table:1}, which also includes the jet ($\rho_{\rm j}=6\times 10^{-30}$~g~cm$^{-3}$) and ejecta ($\rho_{\rm SN}=2.4\times 10^{-23}$~g~cm$^{-3}$) densities, and the SN location ($x=0,y=L_y/5,z=0$) at $t=0$. The interaction distance was set on kiloparsec distances from the SMBH (thus, $R_{\rm j}=100$~pc), chosen to maximise the chance of an SN occurring inside the jet \citepalias[see][]{bosch23}. 

\begin{figure}
\centering
\includegraphics[width=1\linewidth]{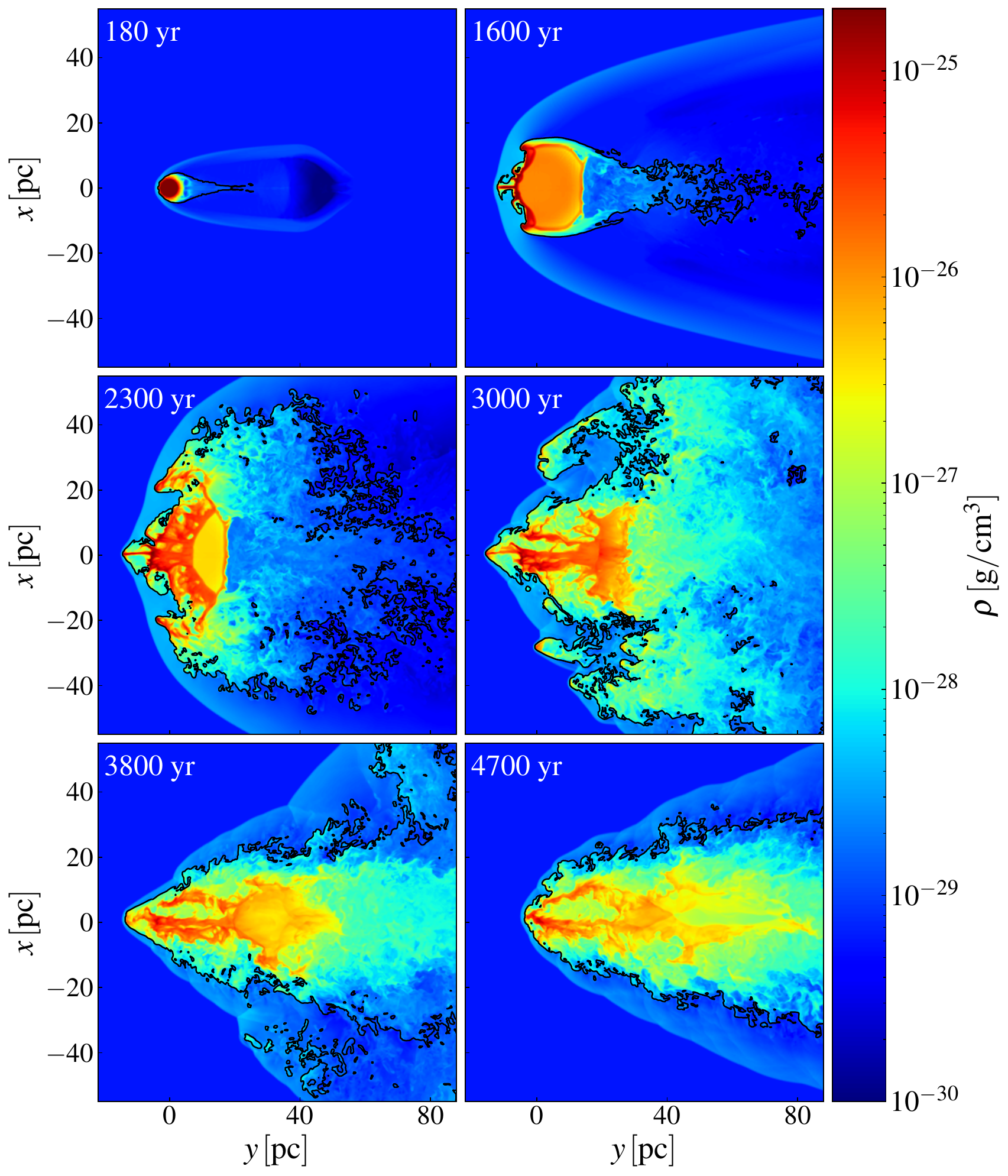}
\caption{Colour maps of the mass density in the $X-Y$ plane at $z=0$ obtained with the 3D RHD simulation described in Sect.~\ref{simu}, for $t=0.18$, 1.6, 2.3, 3, 3.8, and 4.7~kyr. Overlapped over the 2D map is a contour of the tracer ($f=0.4$; see \citetalias{longo25} for details).}
\label{3d_timeframe_rho_hr}
\end{figure}

The results of our simulations show that the SN~ejecta expansion is free for about 1~kyr, but eventually stops in the upstream direction when the jet ram pressure balances that of the ejecta. As seen in Fig.~\ref{3d_timeframe_rho_hr}, from that point onwards, the jet ram pressure drives the evolution of the ejecta, which expands and disrupts on a timescale similar to its sound crossing time. Finally, the ejecta material becomes turbulent and mixes with the jet flow while being driven towards the end of the numerical box. The average jet mass load and jet deceleration are significant: $\sim 25$\% of the jet mass rate during the simulation, and about 40\% of the initial bulk speed when the shock has covered most of the jet section\footnote{This deceleration refers to the whole simulated jet cross-section, as the jet flow almost stops at the locations where it is most strongly shocked by the ejecta.}. The numerical setup used by \citetalias{longo25} was the first to feature a high-resolution grid in 2D\footnote{\cite{vieyro19} had previously presented low-resolution 2D RHD results only, unable to capture in detail the shocked ejecta evolution at the stages considered in \textcolor{blue}{L25}.} and 3D, and the simulated SN ejecta and jet properties were meant to represent a wide range of cases. 

The simulations were run with the hybrid OpenMP+OpenMPI \texttt{Ratpenat} code \citep{perucho10}, which solves the equations of RHD in a conservative form. The code uses high-resolution shock-capturing methods (HRSC), piecewise parabolic methods (PPM) for variables reconstruction, and the Marquina fluxes for the approximate Riemann solver \citep{donat98}. The Synge equation of State \citep{synge57} is implemented to account for leptonic and baryonic populations of particles. 

\begin{table}[t]
\caption{Simulation parameters.}
\label{table:1}
\centering     
\begin{tabular}{|c c c c | }    
\hline

$R_{\rm SN}\,(\rm pc,\rm cells)$ & $L_x,L_y,L_z\,(R_{\rm SN})$ &$E_{\rm SN}\,(\rm erg)$ & $M_{\rm SN}\,(M_\odot)$\\

 1.1, 8  & 100 &$10^{51}$ & 2\\
 \hline
 $R_{\rm j}\,(\rm pc)$& $\rho_{\rm SN}\,(\text{g}{\,}\text{cm}^{-3})$& $T_{\rm SN}\,(K)$ & SN loc. \\
  100 & $2.4\times10^{-23}$ & $10^{9}$ & $L_y/5$\\
\hline
$\rho_{\rm j}\,(\rm g\,\rm cm^{-3})$ &  $L_{\rm j}\,(\rm erg\,s^{-1})$ & $W_{\rm j}$ & $T_{\rm j}\,(K)$ \\
$6\times10^{-30}$ & $10^{44}$ & 2  &  $2\times10^{11}$\\
\hline
\end{tabular}
\end{table} 

\section{Non-thermal model}\label{model}

Particles can be accelerated at the shock crossing the ejecta during the first thousands of years of the interaction, that is, the typical timescale of duration of the shock. These particles remain in the shocked ejecta, while losing energy for a long time because of the slow radiative and adiabatic losses. However, most of the non-thermal energy is expected to be generated in the shock produced in the jet when the obstacle is encountered, because the amount of jet energy dissipated in that shock is much larger than the amount dissipated in the shock crossing the ejecta \citepalias{bosch23}.

\subsection{Electrons energy distribution}

We adopted a phenomenological model for particle acceleration, and just assumed that internal energy in the shocked regions is a proxy of the non-thermal energy, which in slow-cooling particle regimes is a reasonable assumption. We also assumed that diffusive shock acceleration \citep{drury83} is the dominant mechanism in the mildly relativistic shocks of the jet and ejecta. Consequently, we adopted a power-law exponent for the injected particle energy distribution of $p=2$, similar to what is expected in such shocks. We note that a harder (softer) value by $\Delta p$ would harden (soften) the radiation spectrum by $\Delta p/2$ below $\sim$~TeV energies. We computed the emission of electrons (and positrons, if present) via synchrotron and IC scattering; these are the only relevant particles and processes radiation-wise. For the particle energy distributions in the flow frame (in which quantities are primed) in each cell, we adopted a phenomenological approach using a broken power law to account for the effect of synchrotron losses
\begin{align}
\frac{dN'}{dE'} = \left\{\begin{array}{lll}  
K_1 E'^{-1/3} & {\rm if\quad} {\rm 1\,MeV}<E'<E'_{\rm min} \\
K_2 E'^{-p} e^{-E'/E'_{\rm max}} & {\rm if\quad} E'_{\rm min}<E'<E'_{\rm b}\\
K_3 E'^{-(p+1)} e^{-E'/E'_{\rm max}}& {\rm if\quad} E_{\rm b}'<E'<E'_{\rm max}\,,
                              \end{array}
                    \right.
\end{align}
where $E_{\rm max}'$ is the maximum energy of electrons in that cell, $E'_{\rm b}$ the electron energy break in $dN'/dE'$ due to the non-radiative-to-radiative cooling transition, $E'_{\rm min}$ the minimum electron energy associated with the plasma temperature given by the numerical code in each cell (assuming a one-temperature plasma), and $K_1,K_2,K_3$ are constants depending on the non-thermal energy $E'_{\rm NT}$ present in the cell. 
Below $E'_{\rm min}$, given the adopted emissivity $\delta$-function approximations (see below), a power-law index was chosen to mimic the synchrotron emission of electrons with $E'_{\rm min}$ below their characteristic synchrotron frequency (only relevant at the lowest radio frequencies). The global minimum at 1~MeV is equivalent to a Lorentz factor $\gamma \simeq 2$ for electrons.

\begin{figure}
\centering
\includegraphics[width=\linewidth]{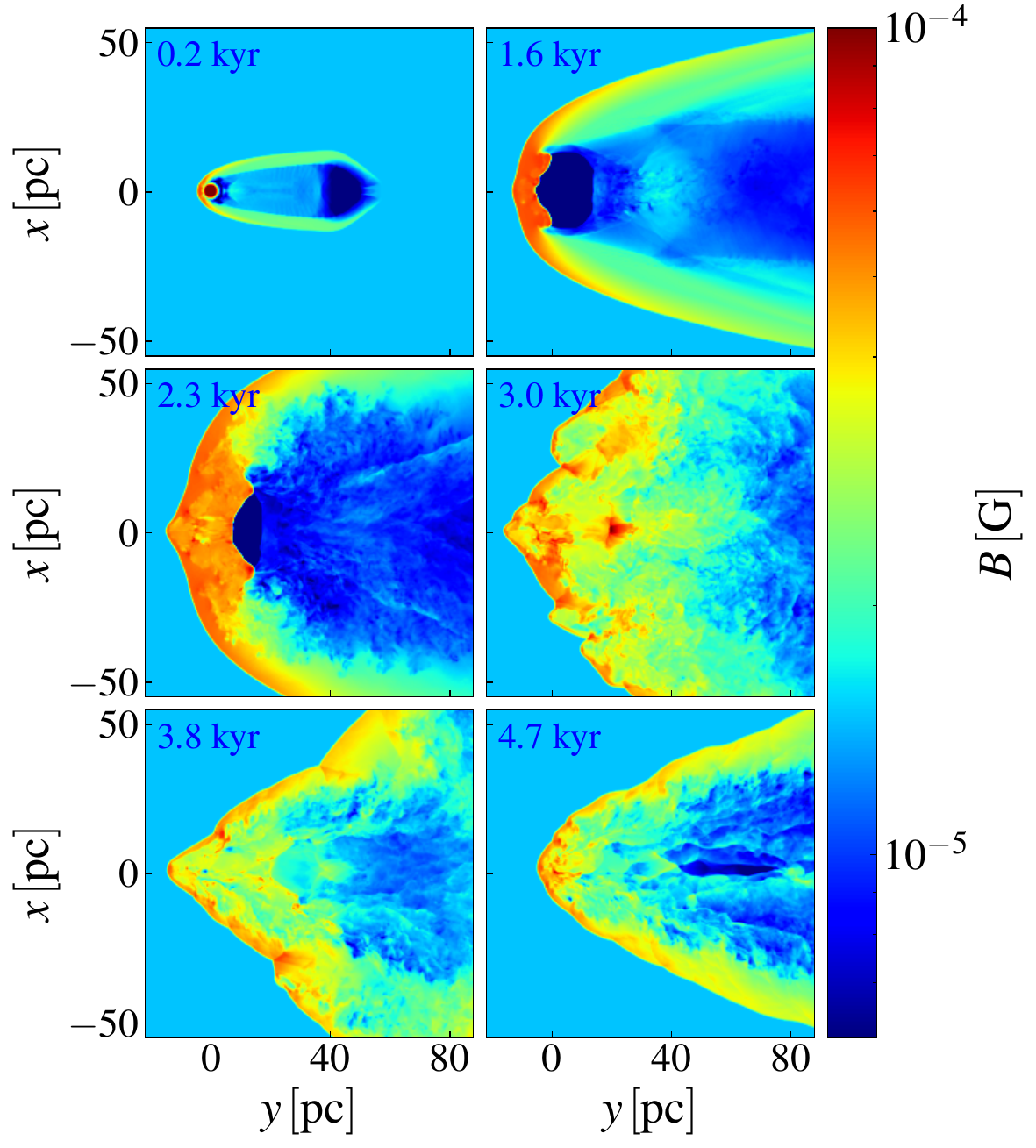}
\caption{Coloured map in 2D of the magnetic field in the lab frame in the $X-Y$ plane at $z=0$, derived with $\eta_{\rm B}=0.01$ for different times of the 3D RHD simulation described in Sect.~\ref{simu}.}
\label{bsnaps}
\end{figure}

For simplicity, $E_{\rm max}'$ for electrons was phenomenologically derived for each cell by balancing the acceleration rate $\dot{E}_{\rm acc} = \eta_{\rm acc}qBc$, where $q$ is the absolute value of the electron charge and $\eta_{\rm acc}$ was fixed to 0.1 (adequate for Bohm diffusion and a mildly relativistic shock), and the synchrotron cooling rate \citep{blumenthal70}
\begin{align}
\dot{E}_{sync} = -\frac{4}{3}\sigma_{\rm T}c\gamma_{\rm max}^{2}U_B=-a_{\rm sync}B^{2}E^{2}_{\rm max}\,.
\label{sync_cooling_rate}
\end{align}
This yields
\begin{align}
    E'_{\rm max} = \frac{1}{W}\sqrt{\frac{\eta_{\rm acc}qc}{a_{\rm sync}B}} \quad {\rm with\quad} a_{\rm sync} = \frac{4}{24\pi}\sigma_{\rm T}\frac{c}{(m_ec^2)^2}\,,
\end{align}
where $W$ is the Lorentz factor of the fluid, $\sigma_{\rm T}$ the Thomson cross section, and $B$ the lab frame magnetic field, derived from taking the magnetic pressure $B'^2/8\pi$ to be a fraction $\eta_B$ of the gas pressure fixed to 0.1 and 0.01 \citepalias[see][for a discussion]{bosch23}. For an injected $B$ perpendicular to the flow direction, one has $B=WB'$. To illustrate the typical $B$-values in different regions, a 2D map of $B$ for $\eta_B=0.01$ and different simulation times is presented in Fig.~\ref{bsnaps}.
Synchrotron losses always dominate over escape from the source (the Hillas limit) and determine the electron $E'_{\rm max}$. In contrast, for protons and nuclei $E'_{\rm max}$ is dominated by escape \citepalias{bosch23}; for these particles and for simplicity, we computed $E'_{\rm max}$ at the location where the particle acceleration efficiency is the highest, adopting the size of the whole interaction region. The break energy $E'_b$ of electrons was found by balancing the non-radiative timescale, that is, the minimum between the source age $t$ and the adiabatic cooling time $t'_{\rm ad}=3/\nabla \cdot ( Wv)$ ($v^{i}$ is the three-vector velocity in the lab frame), and the synchrotron cooling time: $\min[t'_{\rm ad},t] = E_{\rm b}'/|\dot{E}'_{\rm sync}|$. Finally, the distribution normalisation constants were derived as:
\begin{align}
\int_{E'_{\rm min}}^{E'_{\rm max}}E'\frac{dN'}{dE'}dE' = E_{\rm NT}' = \eta V'u\,,
\end{align}
where $u$ is the internal energy density, $V'$ is the flow-frame cell volume, and $\eta$ the internal energy fraction in the non-thermal electrons in the cell, which was fixed here to 0.1 as a reference value; all our non-thermal predictions scale linearly with $\eta$.

\subsection{Spectral energy distribution}

We simplified the derivation of the flow-frame spectral energy distribution for each cell
\begin{align}
    \varepsilon' L'_{\varepsilon'} = \varepsilon' \frac{dN'}{dE'}|\dot{E}'|\frac{dE'}{d\varepsilon'},
\end{align}
by imposing that an electron of energy $E'$ produces a photon of energy $\varepsilon'$, which is a $\delta$-function approximation of the emissivities in which $E'$ and $\varepsilon'$ are linked. The value of $\varepsilon'$ was derived using the characteristic synchrotron photon energy \citep{blumenthal70}
\begin{align}   
\varepsilon'_{\rm sync}=\frac{3hq}{4\pi m_ec}(m_ec^{2})^{-2} B'E'^2,
\end{align}
where $h$ is the Planck constant and $m_e$ is the electron mass when deriving the spectral energy distribution (SED) and $\approx 0.29$ times the latter when deriving the spectral flux and the average scattered photon energy for IC \citep{khangulyan14}. To derive the observer SED, we used the Doppler factor $\delta=1/W(1-\beta\cos\phi)$, with $\beta$ being the emitting flow velocity in $c$ units, and $\phi$ the angle between the flow motion and the line of sight (LOS)
\begin{align}
(\varepsilon L_{\varepsilon})_{\rm obs}=(\delta^{3}/W)\varepsilon'L'_{\varepsilon'}.
\end{align}

The emitting region is transparent to the non-thermal emission, and thus absorption is not taken into account. The total SED was derived by adding the contribution from all the emitting cells, which consisted of those containing shocked jet and ejecta material. Two $\phi$-values were adopted, 20 and 60 degrees, representing a case similar to M87, which leads to moderate boosting, and a case of a more misaligned jet slightly affected by de-boosting, respectively. 

We note that the adopted treatment of particle acceleration, cooling and transport, and the $\delta$-function approximation to compute the SED from each cell, approximately track the energy and spatial distribution of both electrons and non-thermal photons, while immensely reducing the computational costs. These approximations are particularly valid in our case, as the bulk of the non-thermal electrons are mostly dominated by adiabatic losses and source age and evolve smoothly with the background plasma. Moreover, we focused mostly on the power-law region of the particle distribution, more amenable to the use of the $\delta$-function approximation in the calculation of the emissivities. However, it is worth mentioning that the fluxes at the highest synchrotron energies and the IC energies above the peak may be overestimated, as cells far from shocks should in reality have lower electron $E'_{\rm max}$-values due to synchrotron cooling (unless particle re-acceleration is assumed). Finally, the hydrodynamical nature of the simulations presented in Sect.~\ref{simu} does not allow a consistent computation of the magnetic field, and thus we adopted a common approach tying it to the gas pressure. We currently plan to apply the described non-thermal model to the results of a relativistic magnetohydrodynamical simulation performed with the code \texttt{Lostrego} \citep{miralles22}.

\subsection{Synchrotron and inverse Compton}

To compute the synchrotron emission, we followed the procedure just described, adopting the cooling rate from Eq.~(\ref{sync_cooling_rate}), and relating the electron and the photon energies as explained above. To compute the IC emission, we assumed that the target photon fields $\varepsilon_0$ are black bodies with an isotropic photon velocity distribution in the lab frame and transformed them into the flow frame \citep{der02}. We considered target photons of different origins: (i) the infrared (IR) galactic background, with a photon field energy density $u_{\rm IR}=10\,{\rm eV}\,{\rm cm}^{-3}$, reasonably close to the galactic central regions, and an IR photon energy $\varepsilon_{\rm 0,IR}=0.1 \,{\rm eV}$; and (ii) the cosmic microwave background (CMB) with $u_{\rm CMB}\approx 0.25\,{\rm eV}\,{\rm cm}^{-3}$ and $\varepsilon_{0,CMB}\approx 1.6\times10^{-3}\,{\rm eV}$. The thermal and the hadronic radiations (photons and neutrinos) from the whole interaction structure are negligible given the large scales involved.
Regarding the IC cooling rate, we used an expression that accounts for both the Thomson and Klein-Nishina (KN) regimes for a black-body target photon field \citep{bosch09,khangulyan14}\footnote{We adapted the expression in \cite{bosch09} as $\dot{E}'_{\rm IC} =(u_{\rm IR/CMB}'/u_{\rm rad}') m_ec^{2}\dot{\gamma}_{\rm IC}'$, where $u_{\rm rad}=4\sigma/cT^{4}$ and $T=\varepsilon_0/3k_{\rm b}$.}.
We did not take into account the local synchrotron field for the synchrotron self-Compton, which in our calculations has an energy density similar to that in IR but with an overall much smaller IC interaction probability because of a much stronger impact of the KN effect. Emission from jet regions much closer to the SMBH may have been considered as IC targets as well. If it were isotropic and carried 1\% of the jet power, this radiation would have an energy density of about $0.2\,\rm eV\,\rm cm^{-3}$, $\ll u_{\rm IR}$. If more powerful and strongly beamed, this radiation might be significant, but strong KN and IC geometry effects and the  mildly relativistic motion of the shocked jet flow would all tend to diminish the role of such a component (putting the impact of the specific LOS aside, which further complicates the analysis). The role of this inner jet component is not important for hadronic emission processes (e.g., $p\gamma$) either, as the corresponding cross section is much smaller than the IC one. Given all this, we did not include the emission from the inner jet at this stage.

\section{Results}\label{resu}

\begin{figure}
    \centering
    \includegraphics[width=\linewidth]{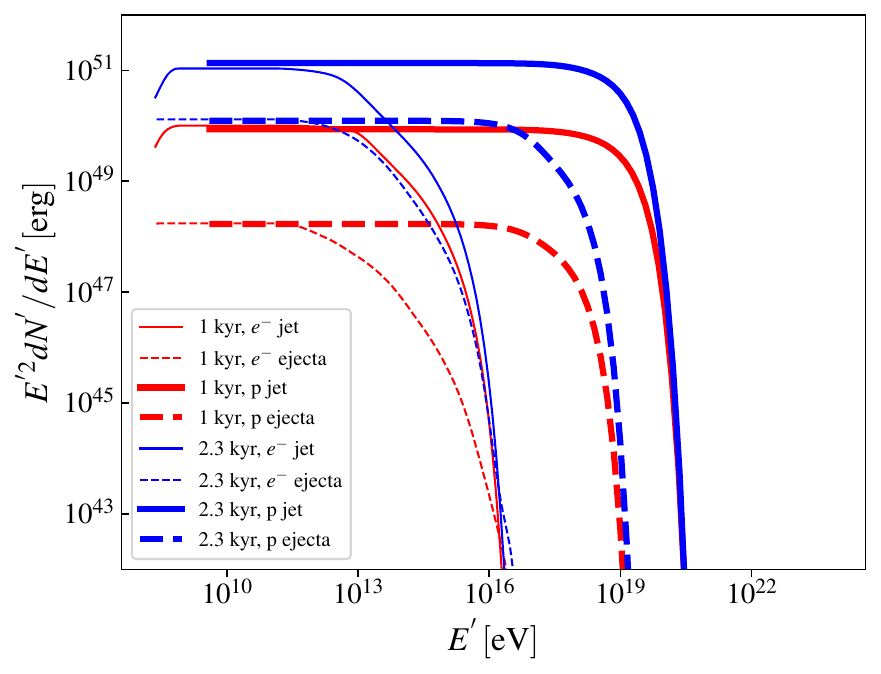}
    \caption{Evolution of energy distributions of the non-thermal populations of electrons and protons in the shocked jet and ejecta regions, for times 1 and 2.3~kyr, and $\eta_B=0.1$. See the legend for details.}
    \label{edist}
\end{figure}

Figure~\ref{edist} shows the energy distributions of non-thermal electrons and protons in the shocked jet and ejecta regions for $t=1$ and 2.3~kyr, and $\eta_B=0.1$. As expected \citepalias{bosch23}, the total non-thermal energetics is about one order of magnitude lower in the ejecta, even when it has acquired large amounts of internal energy due to jet impact, at $t=2.3$~kyr. We adopted $\eta=0.1$ for both electrons and protons, which may not be the case (this choice does not affect our conclusions). The effects of synchrotron losses are clear in the case of electrons, while protons can reach ultra-high energies. We also calculated the emission from radio to $\gamma$-rays produced by the non-thermal electrons in the shocked jet and ejecta. In particular, in our analysis of the temporal evolution and spatial distribution of the emission, we focused on the following reference energy bands: radio ($\sim 10$~GHz); optical ($\sim 5\times 10^{14}$~GHz); X-rays ($\sim 3$~keV); and high-energy and very high-energy $\gamma$-rays (HE $\sim 100$~MeV and VHE $\sim 1$~TeV, respectively). As seen in Fig.~\ref{3d_timeframe_rho_hr}, the bow shock structure encompassing the shocked jet and ejecta flows covers the 3D grid section $X-Z$ (perpendicular to the jet motion) after $\sim 3\times 10^3$~yr. We note that the shocked flow internal energy, relevant for the dominant non-thermal emission, is largely concentrated within the grid on the timescales of interest, and so we neglected the emission from plasma escaping through the grid boundaries. Much larger grid simulations, very difficult to carry out at the same level of detail, would allow the characterisation of larger-scale, more diffuse emission at later times. The jet power, fixed to $L_{\rm j}=10^{44}\,\ergs$, was chosen such that the number of objects in the nearby Universe is not negligible, that is, there would be too few of them when more powerful, and they would be too weak as sources otherwise \citepalias[as discussed in][]{bosch23}. We adopted a distance similar to that of M87, $z\approx0.0035$/$15$~Mpc, taken as a reference value to compare with the sensitivity levels similar to those of some detectors operating in radio (Very Large Array, VLA), optical (\textit{Hubble} Space Telescope, HST), soft X-rays (\textit{Chandra}), and HE (\textit{Fermi}) and VHE (the forthcoming Cherenkov Telescope Array Observatory, CTAO). This comparison is semi-quantitative and aimes only to illustrate the observational potential of these interactions; more accurate observational predictions are left for future work.

\subsection{Spectral energy distribution}

\begin{figure}
    \includegraphics[width=\linewidth]{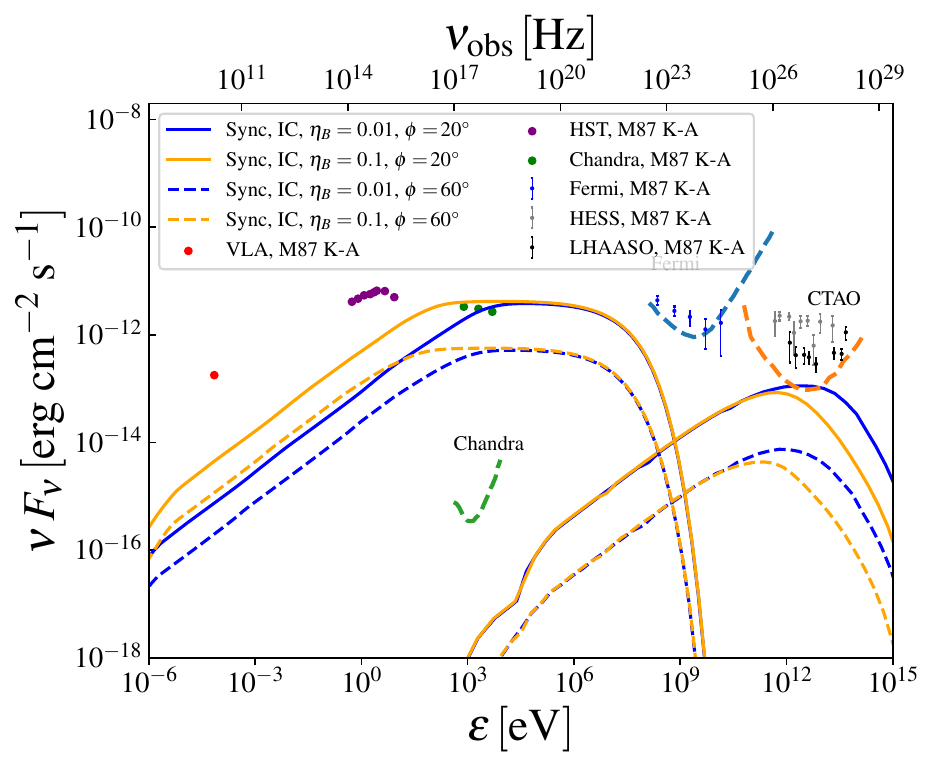}
    \caption{Spectral energy distributions featuring synchrotron and inverse Compton emission for $z\approx0.0035$/$15$~Mpc, $\phi=20^\circ$ (solid lines) and $60^\circ$ (dashed lines), and $\eta_B=0.01$ (blue lines) and $0.1$ (orange lines). Sensitivity curves are plotted for \textit{Chandra} (see the {\it Chandra} \href{https://cxc.cfa.harvard.edu/csc1/sens/index.html}{web page}), \textit{Fermi} \citep{fermi12} and CTAO \citep{ctao19}. Observational data on M87, and only its knot~A in radio, optical and X-rays (see Sect.~\ref{disc}), is given for the VLA and \textit{Hubble} Space Telescope (red and purple dots, respectively; \citealt{perlman01}), \textit{Chandra} (green dots, \citealt{he26}), \textit{Fermi} (blue dots, \citealt{zhang18}), High Energy Stereoscopic System (grey dots, \citealt{magic20})) and the Large High Altitude Air Shower Observatory (black dots, \citealt{Cao24}).}
    \label{sed}
\end{figure}

Figure~\ref{sed} shows the synchrotron and IC SEDs for the whole shocked jet and ejecta regions for $\phi=20^\circ$ and $60^\circ$, and $\eta_B=0.01$ and 0.1, at $t=2.3$~kyr. The ejecta contribution is negligible and is therefore not shown. The IC emission is fainter than the synchrotron emission by more than one order of magnitude, and peaks around 1~TeV due to the dominant IR target photons of IC, reaching fluxes $\approx 10^{-13}$~$\eflux$ for both $\eta_B=0.01$ and $0.1$ and $\phi=20^\circ$. We note that a substantially brighter galactic central region would lead to a significantly higher IC flux, as this is $\propto u_{\rm IR}$. On the other hand, the synchrotron component peaks in X-rays, with fluxes somewhat below $\sim 10^{-11}$~$\eflux$ again for both $\eta_B=0.01$ and $\eta_B=0.1$ and $\phi=20^\circ$. Doppler (de)boosting leads to a difference in flux by a factor of $\sim 10$ between $\phi=20^\circ$ and $60^\circ$.
Increasing $\eta_B$, associated with an enhancement of synchrotron losses, leads to a softening in the synchrotron and IC spectra at their intermediate energies as $E'_{\rm b}$ decreases, with radio and optical fluxes (synchrotron) growing and the VHE one (IC) decreasing; for the adopted $p=2$, at those energies the synchrotron SED goes from being $\propto\varepsilon^{1/2}$ to a flat slope, while the IC SED steepens further due to the Thomson-to-KN regime transition. The predicted total radio, optical and X-ray fluxes at the reference distance of $z\approx0.0035$/$15$~Mpc can be detectable for facilities such as the VLA and HST (see below) and {\it Chandra}. In $\gamma$-rays, IC emission is far from the sensitivity curve of {\it Fermi} at HE, but may be detectable at VHE by CTAO, in particular for host galaxies with high $u_{\rm IR}$ in their central kiloparsec region.

\subsection{Light curves}

\begin{figure}
    \includegraphics[width=\linewidth]{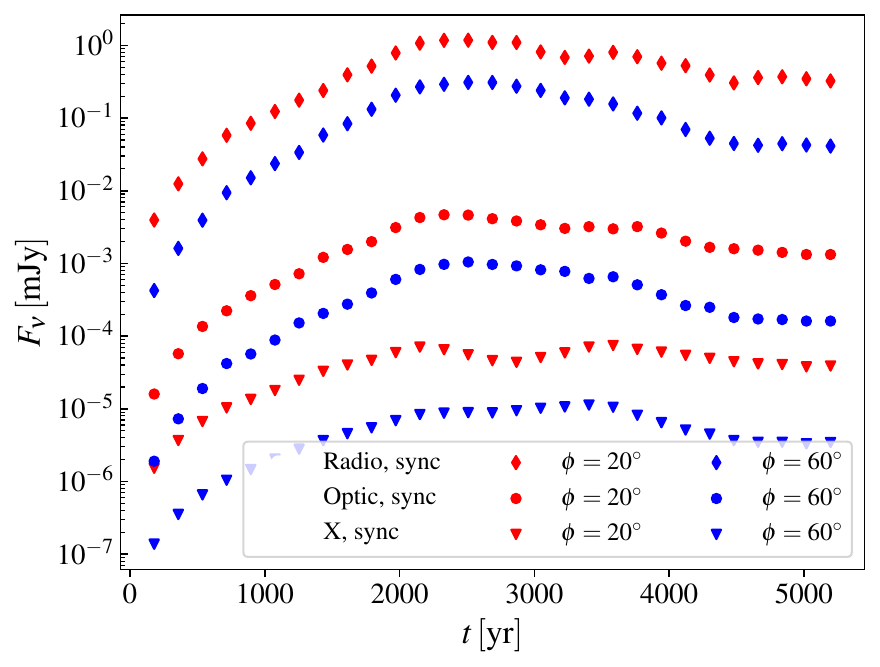}
    \caption{Light curves of the spectral fluxes in mJy at $z\approx0.0035$/$15$~Mpc normalised accounting for typical instrument angular resolutions (see text for details), for $\phi=20^\circ$ (red) and $60^\circ$ (blue), and $\eta_B=0.1$, for the synchrotron emission in radio ($10\,{\rm GHz}$, diamonds), optical ($5\times 10^{14}\,{\rm Hz}$, dots) and $X$-rays ($3\,{\rm keV}$, triangles). We note that the representation now is different from  that of Fig.~\ref{sed} (spectral flux versus SED, respectively), so for instance the radio emission, the least luminous energy band there, appears at the top here.}
    \label{lelc}
\end{figure}

\begin{figure}
    \includegraphics[width=\linewidth]{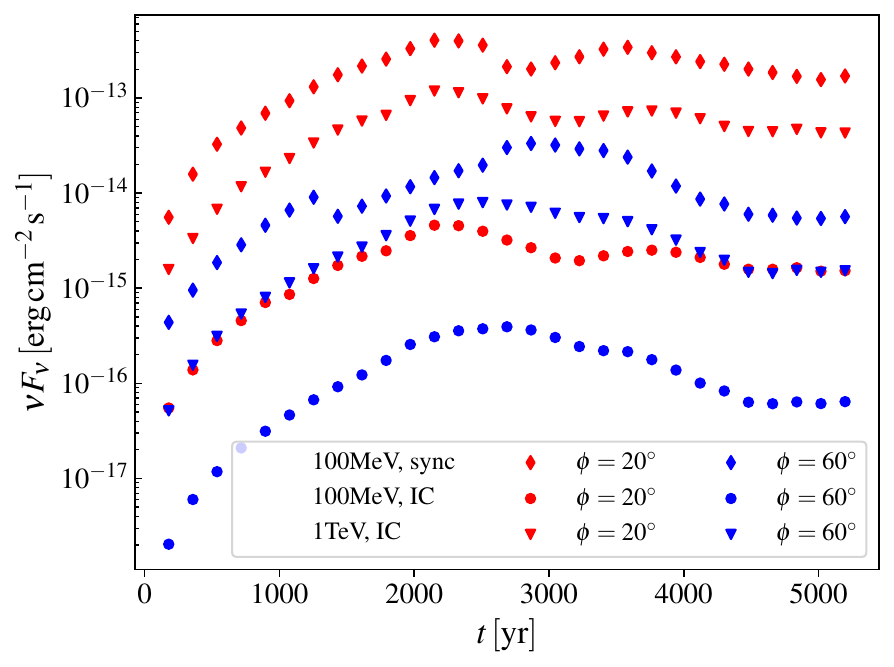}
    \caption{Light curves of the total SED fluxes at $z\approx0.0035$/$ 15$~Mpc, for $\phi=20^\circ$ (red) and $60^\circ$ (blue), and $\eta_B=0.01$, at $100\,{\rm MeV}$ (synchrotron; diamonds), $100\,{\rm MeV}$ (IC; dots) and $1\,{\rm TeV}$ (IC; triangles).}
    \label{helc}
\end{figure}

The light curves (LCs) of the spectral fluxes in radio, optical and X-rays, computed up to $\approx 5200\,{\rm yr}$ and fixing $\eta_B$ to $0.1$ are shown in Fig.~\ref{lelc}, while the LCs of the SED fluxes at HE and VHE for $\eta_B=0.01$, also up to $\approx 5200\,{\rm yr}$, are presented in Fig.~\ref{helc}. Radio, optical, and X-ray LCs were normalised dividing the total fluxes by the approximate angular surface of the whole emitting region, $\sim 1\times 1$~arcsec$^{2}$, and multiplying by an angular surface associated with the instrument resolution in each band, $S_{\rm radio/opt}\sim \pi \,0.1^2$~arcsec$^{2}$ in radio and optical, and $S_{\rm X}\sim \pi\, 0.25^2$~arcsec$^{2}$ in X-rays. All LCs peak around 2.5~kyr and become roughly steady, showing some modulation on a timescale of 0.1--1~kyr due of the complex hydrodynamical evolution of the simulated shocked flows. With the reduction of the jet shock surface at $t\gtrsim 4$~kyr, a drop in the internal energy of the shocked flow occurs, and as the latter leaves the grid, the emission levels start to slowly decrease. At this point, to track the evolution for significantly longer times, calculations that include a much larger region (at least of size $\sim$~kpc) would be required. The shape of all the synchrotron and IC LCs is similar, with any visible small difference being due to the magnetic and the IC target photon fields evolving differently with time; the former depends on the local energy density evolution and the latter is constant. The synchrotron LCs shown in the figures are produced by electrons in different regions of their energy distribution, that is, below (radio and optical) and above (X-rays and 100~MeV) with respect to $E'_{\rm b}$. Thus, any present small difference between the groups of LCs is related to that fact. Nevertheless, only longer, larger-scale RHD simulations would be able to derive more distinctive patterns of evolution for the IC and synchrotron emission at different energy bands.

\subsection{Emission maps}

We computed 2D colour maps of the LOS-integrated emission for $t=1$ (early), 2.3 (peak), and 3.8~kyr (late), $\phi=20^\circ$ and $60^\circ$, $\eta_B=0.1$ and $z\approx0.0035$/$ 15$~Mpc, in the following energy bands: radio, shown in Fig.~\ref{2dradio}; optical, in Fig.~\ref{2doptic}; X-rays, in Fig.~\ref{2dX}; and HE (100~MeV) and VHE (1~TeV), in Figs.~\ref{2D100mev} and \ref{2D1tev}, respectively. The flux in radio, optical and X-rays at each map point ($F_{\rm map}$) corresponds to the flux at each specific energy, produced by a LOS-column of emitting plasma with cross section $S_{\rm pixel}=2\times 2$~milliarcsec$^2$ ($F_{2{\rm mas}\times2{\rm mas}}$), and normalised to a reference angular surface as: 
\begin{align}  
F_{\rm map}=F_{2{\rm mas}\times2{\rm mas}}\frac{S_{\rm radio/opt/X}}{S_{\rm pixel}}\,. 
\end{align}
The units adopted are mJy in radio, nJy in optical, and SED flux units (i.e. erg~cm$^{-2}$~s$^{-1}$) in X-rays. We also smoothed the distribution of $F_{\rm map}$ in each band with Gaussians of $\sigma_{\rm Gauss}=0.1$ (radio and optical), and 0.25 (X-rays); the corresponding non-smoothed 2D maps are presented in Figs.~\ref{2dradio_ng}, \ref{2doptic_ng} and \ref{2dX_ng}. We note that the normalisation to derive $F_{\rm map}$ for radio, optical and X-rays was chosen to illustrate the approximated flux produced from any particular source region with a size similar to a fiducial instrument angular resolution. The approximate radio (e.g., VLA) and optical (e.g., HST) detectability limits would be of the order of magnitude 0.01~mJy and 100~nJy, respectively, while the sensitivity curve of \textit{Chandra} is shown in Fig.~\ref{sed}.
For the 2D maps at HE and VHE, no smoothing was applied because the source size is point-like for todays's instrumentation. At these energies, $F_{\rm map}$ was first derived in SED flux units, and then normalised to the peak of the emission. 

The 2D maps clearly show morphological differences when comparing $\phi=20^\circ$ (left panels in figures) and $60^\circ$ (right panels), with a much more flattened shape in the latter case, for which fluxes are also lower due to Doppler (de)boosting, and significantly more concentrated towards the jet shock region. Fixing $\phi$, the distribution of the emission in the sky in all bands is similar. However, the emission is stronger in the regions of high non-thermal energy density (synchrotron and IC) and strong magnetic field (synchrotron).
Many of the 2D maps feature limb-brightening effects, with the central regions appearing slightly fainter, as a result of the LOS integration, and the fact that the ejecta is a much weaker emitter\footnote{We remind that the emitter is optically thin in all the shown energy bands, as magnetic, radiation and matter fields are very diluted in the large studied region.}.

\section{Summary and discussion}\label{disc}

Based on the results obtained by \citetalias{longo25} of a 3D RHD simulation of the interaction of an SN ejecta with an extragalactic jet, and adopting a semi-phenomenological but physically informed model for particle acceleration, transport, and energy evolution, we computed the SEDs, LCs and sky-projection 2D maps of the non-thermal emission produced by this interaction from radio to VHE. The jet power ($10^{44}$~erg~s$^{-1}$), ejecta mass ($2\,$M$_\odot$) and interaction jet-scale ($\sim$kpc) were chosen to represent the most common events of this kind, whereas the parameters of the non-thermal model and the magnetic field were fixed to standard values to derive results not directed towards explaining a particular source, but which could provide an illustration of the potential non-thermal outcome of SN~ejecta--jet interactions. 

For a source with a non-thermal electron population containing $\sim 10$\% of the shocked flow internal energy, an efficient acceleration in the Bohm regime in a mildly relativistic shock, and the magnetic pressure being 1 and 10\% of the gas pressure, we found that the levels of radio, optical and X-ray radiation could be detectable at distances $\lesssim 100$~Mpc for typical instrument sensitivities. For typical angular resolutions of instruments operating at those energies, the structure may be resolved at the adopted fiducial distance of $z\approx0.0035$/$ 15$~Mpc (and potentially much further if using milliarcsec angular resolution instruments with enough sensitivity). At both HE and VHE the simulated region should appear point-like.
At VHE energies, the interaction structure fluxes may be within the reach of the sensitivity of CTAO, in particular if the galactic central kiloparsec region is bright. At HE, the emission seems hard to detect for current facilities operating at those energies, which is aggravated by the fact that we may be overestimating the fluxes at the highest synchrotron energies. However, we emphasise that our simplified approach is realistic in the remaining energy bands, at least at a semi-quantitative level, as the associated radiation is not significantly affected by the specific $E_{\rm max}$-value of electrons.  

It is worth noting that the simulation providing the dynamical basis for our non-thermal calculations assumed a stationary jet, although jets of active galactic nuclei present variability in very different scales. The fact that the interaction scales are $\sim 100$~pc implies that any jet property variability on timescales shorter than several hundreds of years, such as, for instance, the year-scale variability found by very long baseline interferometry observations, will be smoothed out. Nevertheless, variability on 100~yr timescales or even shorter may be coupled with instability growth within the interaction region, making the overall evolution of the shocked flows even more chaotic and turbulent. Although the interaction process is already very disruptive in our simulation, stronger turbulence could enhance the role of the Fermi~II acceleration mechanism (\citealt{fer49}; see also, e.g., \citealt{rie07} for a discussion of diffusive acceleration processes in jets), leading to more distributed acceleration in the shocked jet flow. This would likely produce a moderately higher emission on all bands, less confined to the strongest shocks and extended over a larger region. Unfortunately, given that our acceleration model is mostly phenomenological (as explained in Sect.~\ref{model}), it is not possible to quantify the impact of this effect at this stage, but it deserves a devoted study that includes a more physical and nuanced acceleration description.

The SN~ejecta--jet interaction scenario was applied to knot~A of M87 by \cite{he26} \citep[see][for an early discussion]{blkosnr} using a simple analytical dynamical model based on \citetalias{bosch23}, and a purely phenomenological one-zone radiation model. The observational data and their computed SEDs can be found in fig.~2 of \cite{he26}; we also show in Fig.~\ref{sed} the data from that work to give context to the present discussion. We emphasise that our calculations were not aimed at explaining knot~A in M87 in particular, which would require a much lighter calculation apparatus more suitable for data fitting analysis, as done in \cite{he26}, but rather at the typical non-thermal luminosities at different energy bands using a simple non-thermal particle model. Nevertheless, a qualitative comparison is in order: as seen in Fig.~\ref{sed}, the X-ray luminosities obtained here could easily reach the observed values, but not the observed spectrum, which is softer (which can impact cosmic ray estimates; see below). On the other hand, the predicted VHE emission is about one order of magnitude below the VHE emission from the whole radio galaxy, although it is worth recalling that knot~A cannot be disentangled from the rest of the source at these energies, so its actual VHE luminosity is unknown. In radio and optical, the predicted fluxes are below those seen in knot~A, although there is an important issue to point out at those frequencies: in \cite{he26}, the authors added an ad hoc additional lower-energy electron population to explain that emission, whereas here we assumed the presence of only one power law-like population of non-thermal electrons accelerated in the jet and ejecta shocks. However, a relic population of electrons of moderate energies ($\ \sim 1-100$~GeV), and softer than $\sim E^{-2}$ to explain the radio and optical fluxes together, could already be present in the jet upstream from the interaction structure, which would easily increase the radio-to-optical synchrotron emission by one order of magnitude due to both adiabatic compression and magnetic field enhancement at the jet shock. Therefore, the synchrotron emission from a relic electron population, accelerated closer to the SMBH and faint (or undetectable) before the jet shock, would become much brighter at the interaction region. This would imply an increase in energetics by a factor of several with respect to our simpler model for those electrons emitting in radio and the optical via synchrotron, which can be accommodated in the proposed scenario given the adopted non-thermal fraction $\eta=0.1$.

Finally, since the structure of the SN~ejecta--jet interaction shows prominent mixing (see, e.g. \citepalias{longo25}, protons from both the jet and the ejecta, and large quantities of heavier nuclei from the latter, can potentially be accelerated  up to $\sim 10\eta_BZ$~EeV under the conditions assumed in the present work \citepalias[$Z$ being the atomic number; see, e.g.,][for details]{bosch23}. In particular, matter and radiation interactions should be inefficient cooling channels for these particles \citepalias{bosch23}. Thus, ultra-high-energy cosmic rays (UHECR) accelerated in this interaction, and capable of escaping the jet before suffering adiabatic losses, could reach the jet cocoon outside the host galaxy. Given the slow adiabatic and hadronic radiation cooling in the jet cocoon, some of these UHECR could eventually reach Earth if originating from a source in the local universe. Under the mentioned regime of inefficient hadronic processes, neutrino production is expected to be negligible on the jet scales considered. Nevertheless, an interaction with a much denser SN~ejecta at the jet base might be a significant neutrino emitter, although the small volumes involved would imply very low-occurrence rates. 
However, a note of caution must be added however if knot~A of M87 were indeed the result of an SN~ejecta--jet interaction, as the soft X-ray spectrum of this knot would suggest that protons and nuclei carry little energy at the highest energies if both leptons and baryons have the same injection energy distribution. Although it is debatable whether this should necessarily be the case, it must be taken into account when assessing the hadronic energetics in knot~A.

\begin{acknowledgements}

This work has received financial support from the Spanish Ministry of Science and Innovation under grants 
PID2022-136828NB-C41/AEI/10.13039/501100011033/ERDF/EU,
PID2022-136828NB-C43/AEI/10.13039/501100011033/ERDF/EU, 
PID2025-168247NB-C41/AEI/10.13039/501100011033/ERDF/UE, and 
PID2025-168247NB-C43/AEI/10.13039/501100011033/ERDF/UE, and the María de Maeztu award to the ICCUB CEX2024-001451-M. V.B-R. is Correspondent Researcher of CONICET, Argentina, at the IAR.
\end{acknowledgements}

\bibliographystyle{aa}
\bibliography{biblio}

\appendix

\section{Two dimensional maps of emission}\label{2dmaps}

In this Appendix we compile the two dimensional maps of emission, given two angles on the LOS, $\phi=20^\circ$ and $\phi=60^\circ$, and five different wavelengths: radio ($10^{10}\,$GHz, ), optical ($5\times10^{14}\,$GHz, ), X-rays (3~keV), high energy (100~MeV, ) and very high energy (1~TeV, ).

\begin{figure}[ht]
\centering
    \includegraphics[width=\linewidth]{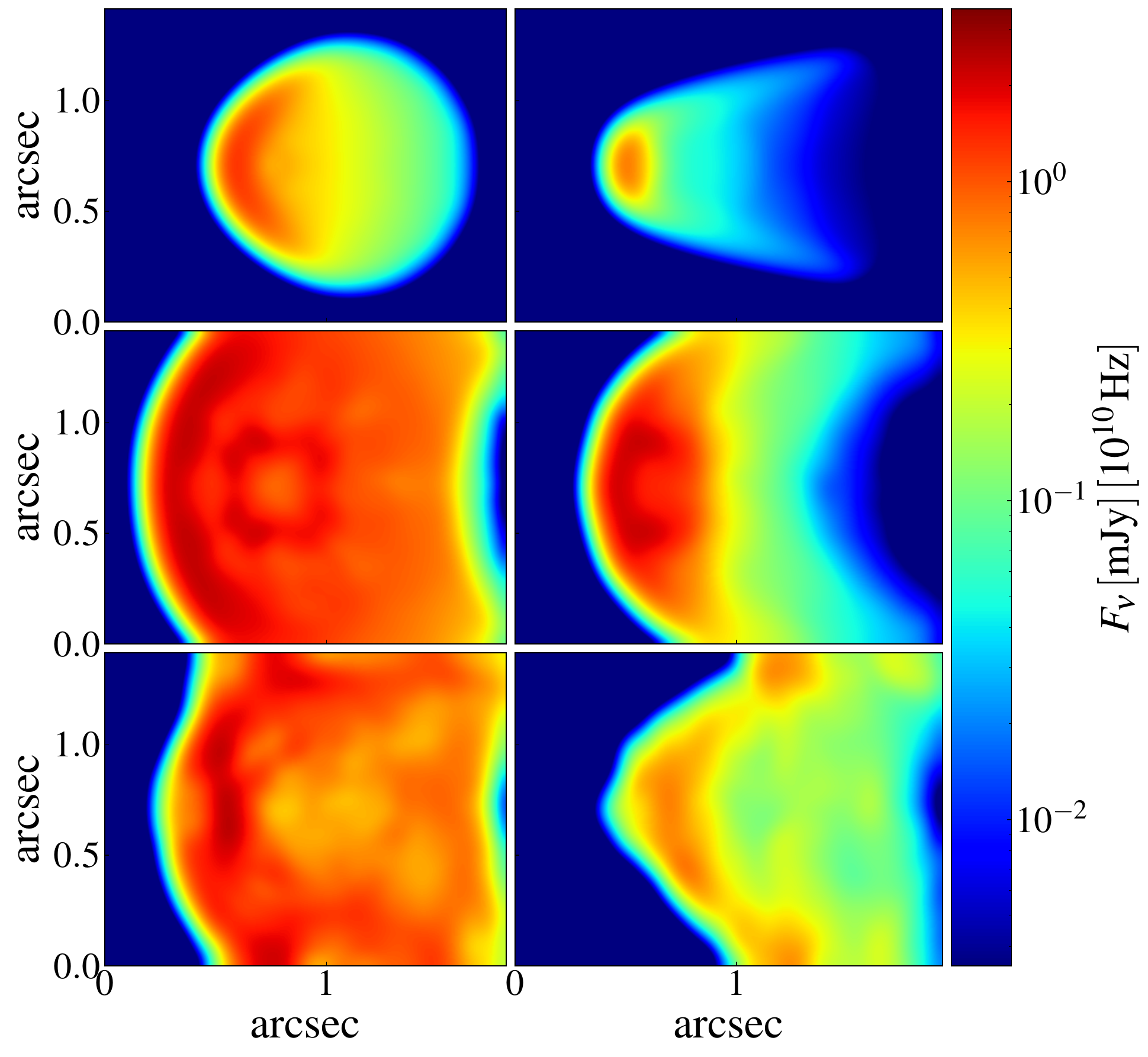}
    \caption{Colour 2D maps of the fluxes at 10~GHz in mJy produced within an angular surface associated to an instrument angular resolution of $\sim 0.2$~arcsec, for $\phi=20^\circ$ (left column) and $60^\circ$ (right column), $z\approx 0.0035$/$ 15$~Mpc, and simulation times 1 (top panel), 2.3 (middle panel), and 3.8~kyr (bottom panel). A Gaussian smoothening with $\sigma_{\rm Gauss}=0.1$~arcsec was applied.}
    \label{2dradio}
\end{figure}

\begin{figure}[ht]
\centering
    \includegraphics[width=\linewidth]{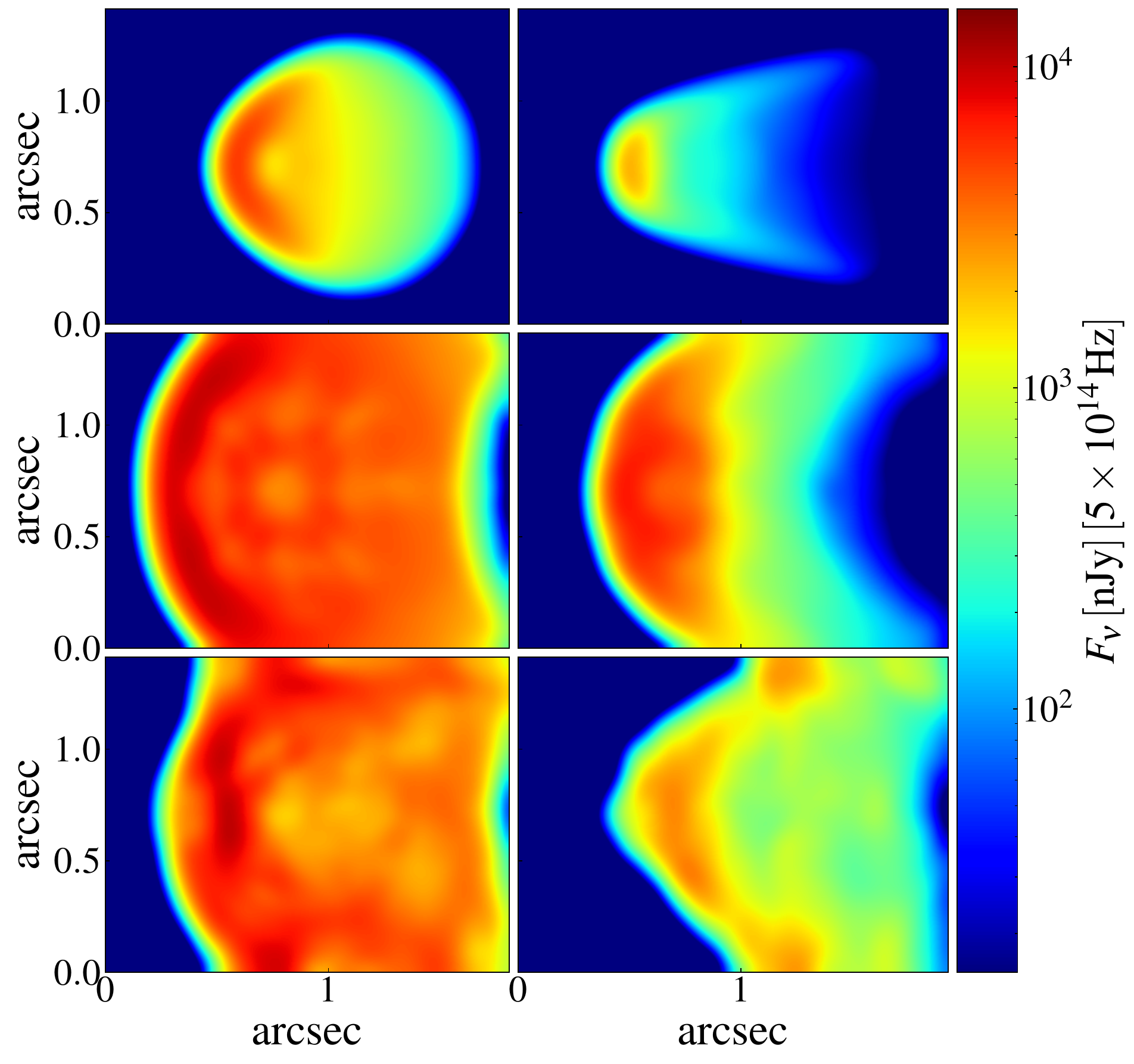}
    \caption{Same as in Fig.~\ref{2dradio}, but at $5\times 10^{14}$~Hz and using nJy units, while keeping the same angular resolution and Gaussian smoothening.}
    \label{2doptic}
\end{figure}

\begin{figure}[ht]
\centering
    \includegraphics[width=\linewidth]{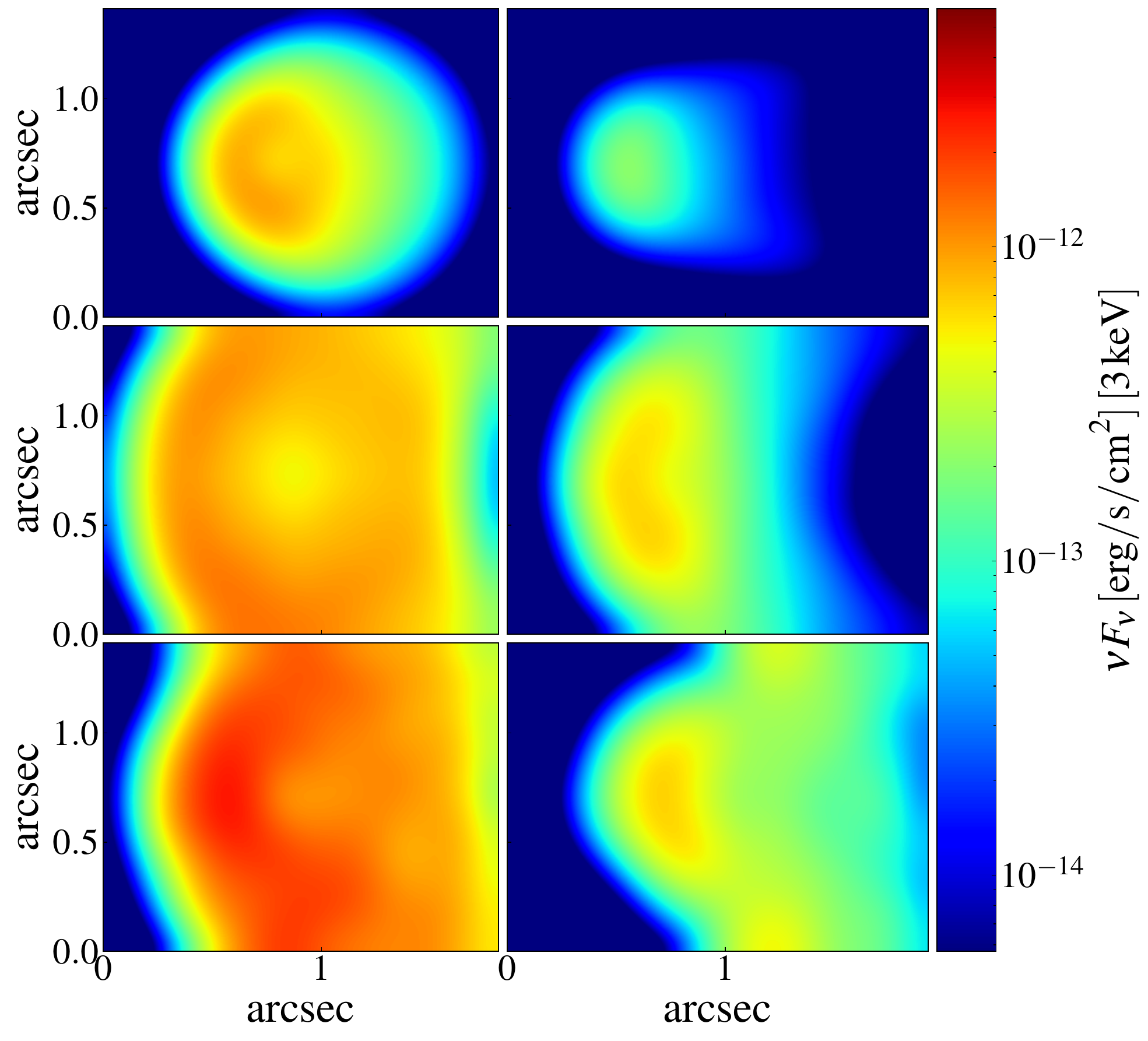}
    \caption{Same as in Fig.~\ref{2dradio} but for the SED flux at 3~keV, and adopting an angular resolution of $\sim 0.5$~arcsec, and a Gaussian smoothening with $\sigma_{\rm Gauss}=0.25$~arcsec.}
    \label{2dX}
\end{figure}

\begin{figure}[ht]
\centering
    \includegraphics[width=\linewidth]{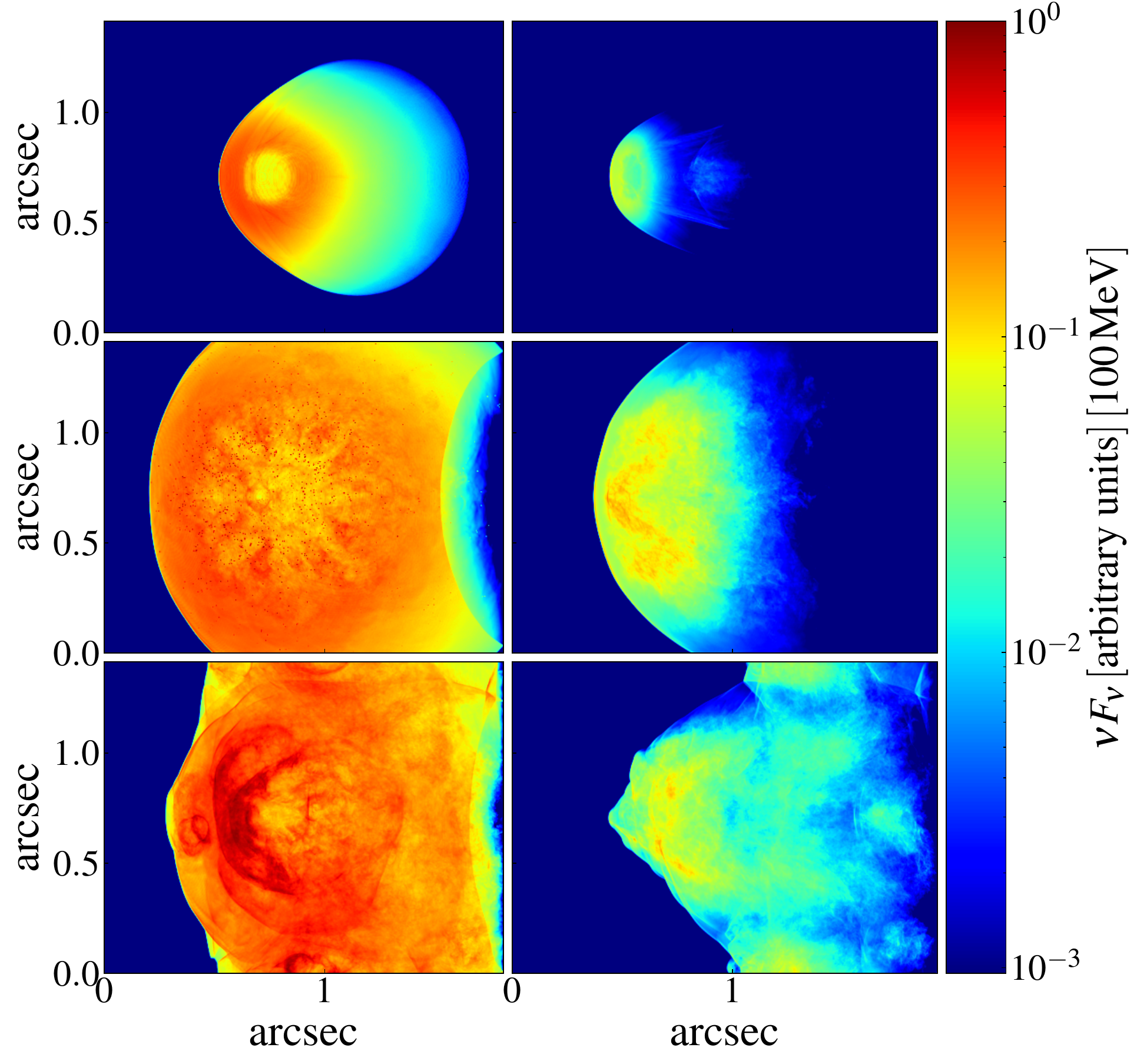}
    \caption{Same as in Fig.~\ref{2dradio} but for the SED flux at 100~MeV in units normalised to the peak of the emission ($3\times10^{-18}\,\eflux$, and without Gaussian smoothening.}
    \label{2D100mev}
\end{figure}

\begin{figure}[ht]
\centering
    \includegraphics[width=\linewidth]{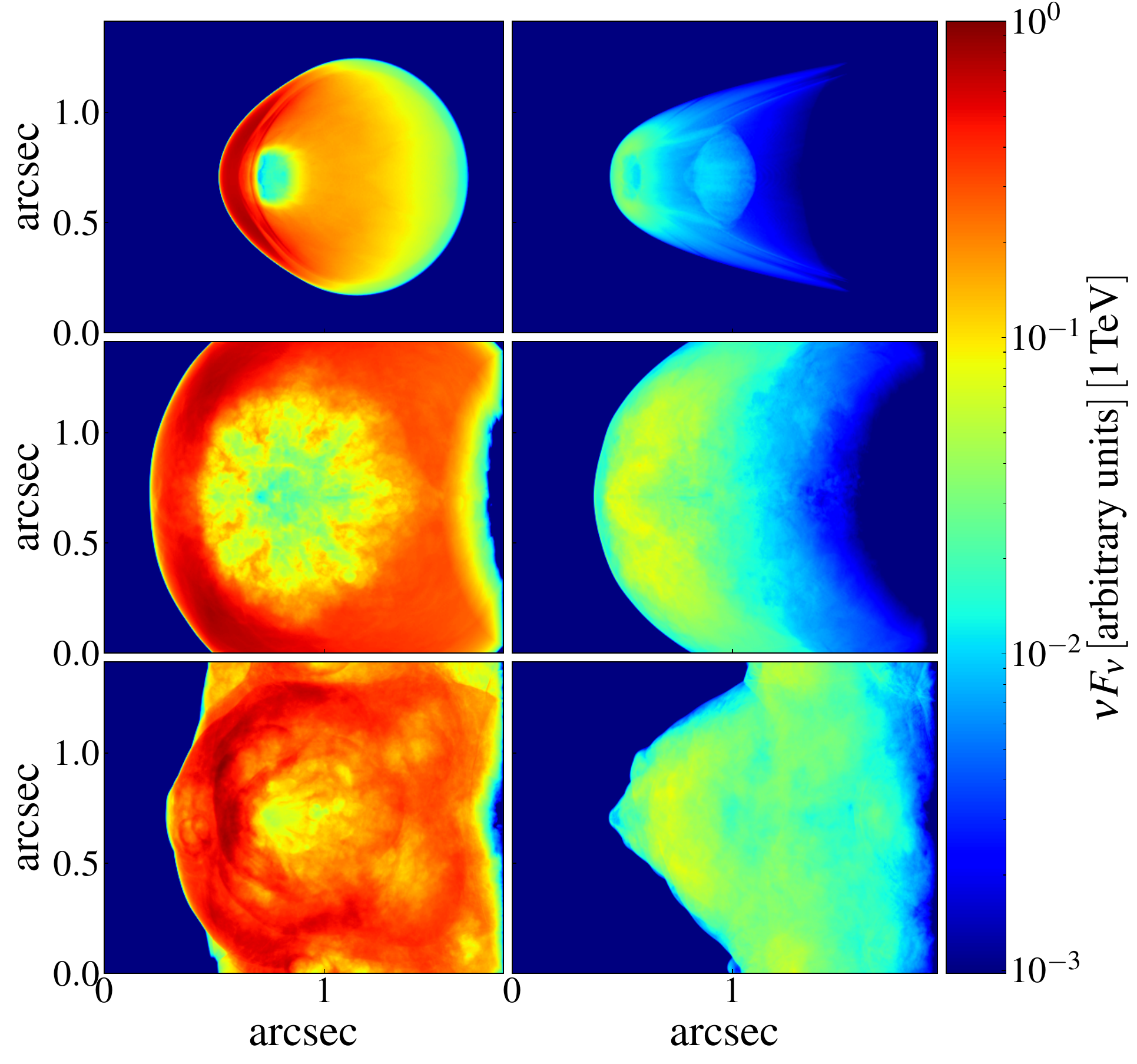}
    \caption{Same as in Fig.~\ref{2D100mev} but at 1~TeV. Peak of emission: $6\times10^{-19}\,\eflux$.}
    \label{2D1tev}
\end{figure}

\begin{figure}[ht]
\centering
   \includegraphics[width=\linewidth]{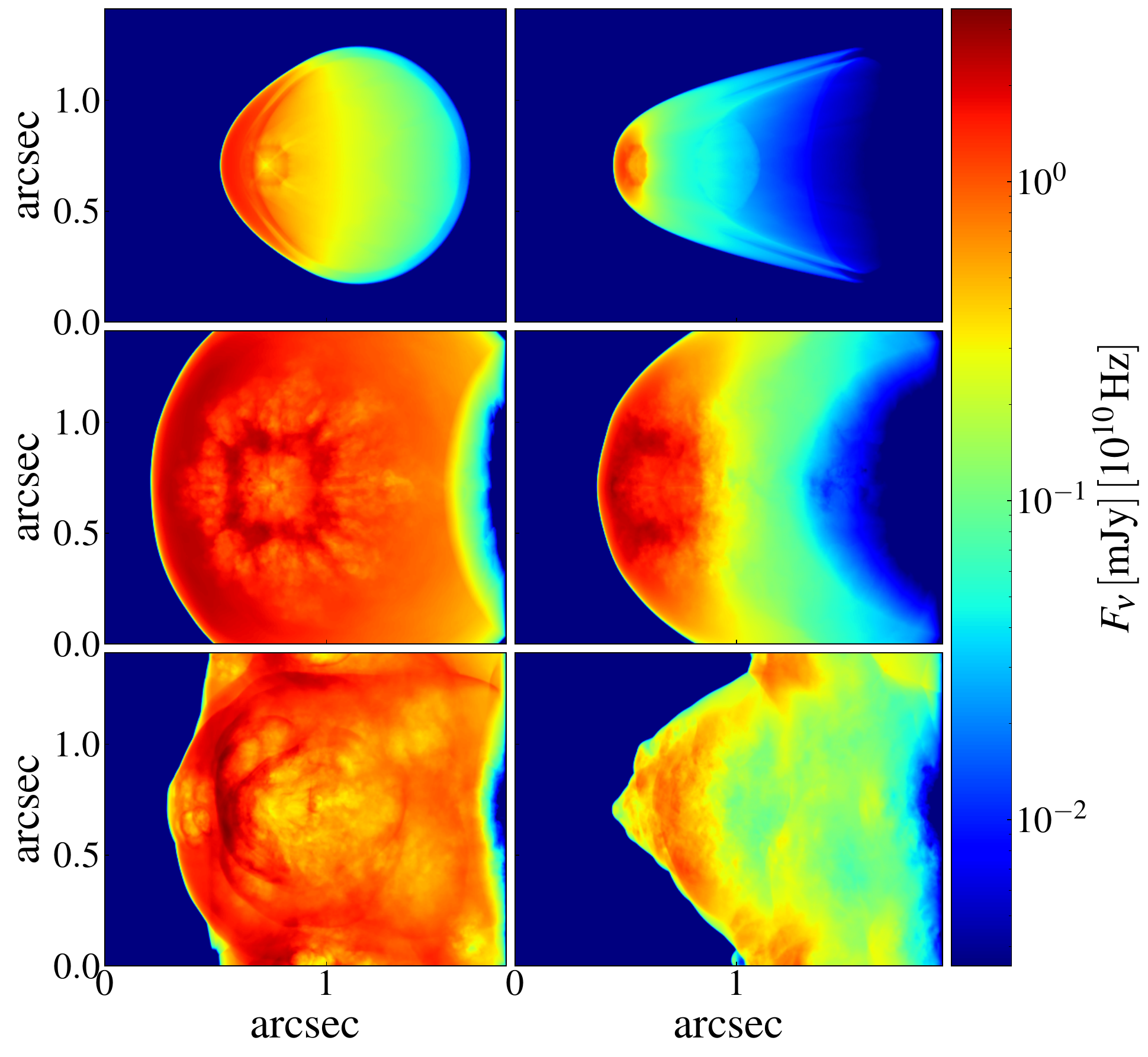}
    \caption{Same as in Fig.~\ref{2dradio} but without Gaussian smoothening.}
    \label{2dradio_ng}
\end{figure}

\begin{figure}[ht]
\centering
    \includegraphics[width=\linewidth]{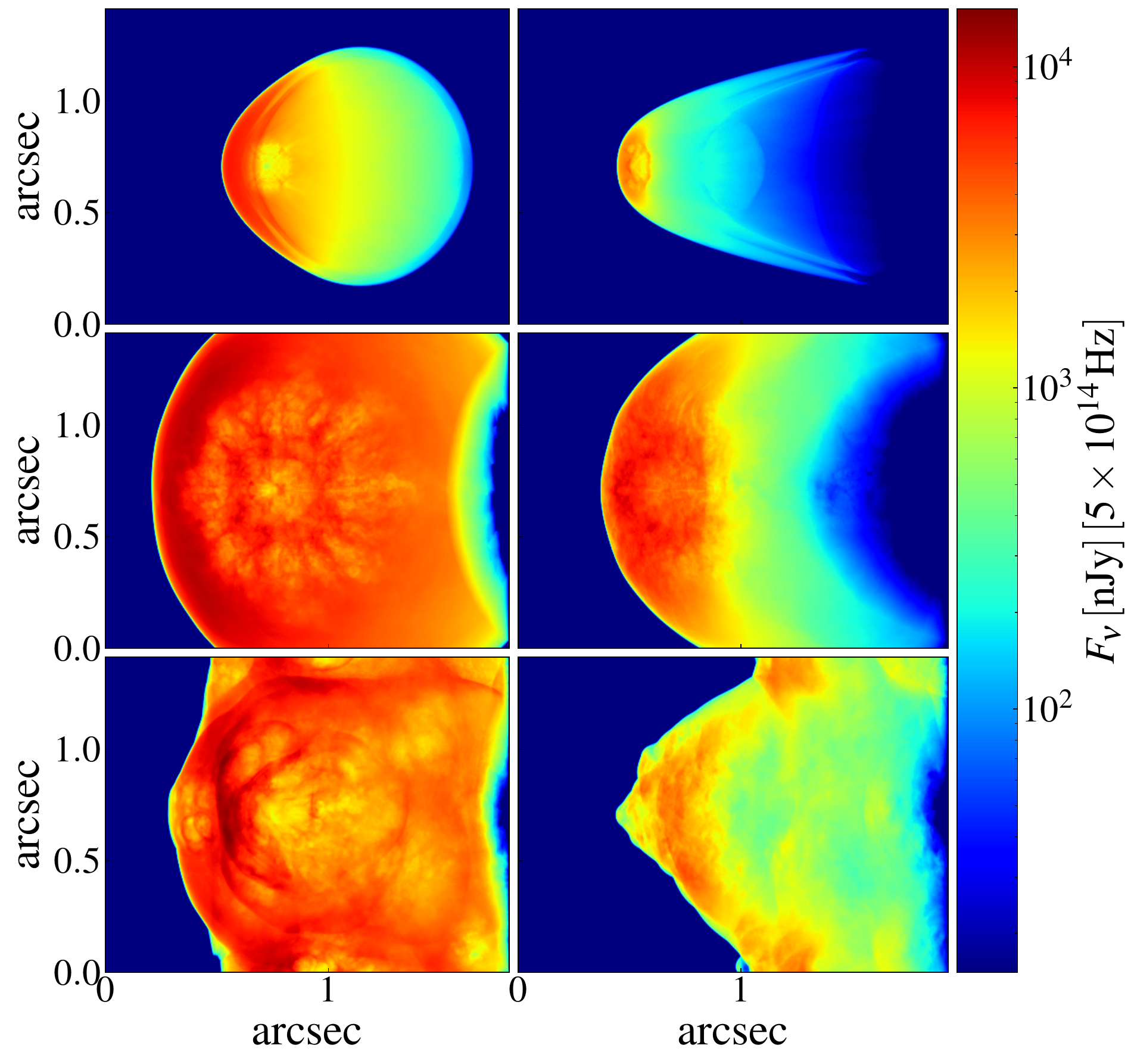}  
    \caption{Same as in Fig.~\ref{2doptic} but without Gaussian smoothening.}
    \label{2doptic_ng}
\end{figure}

\begin{figure}[ht]
\centering
    \includegraphics[width=\linewidth]{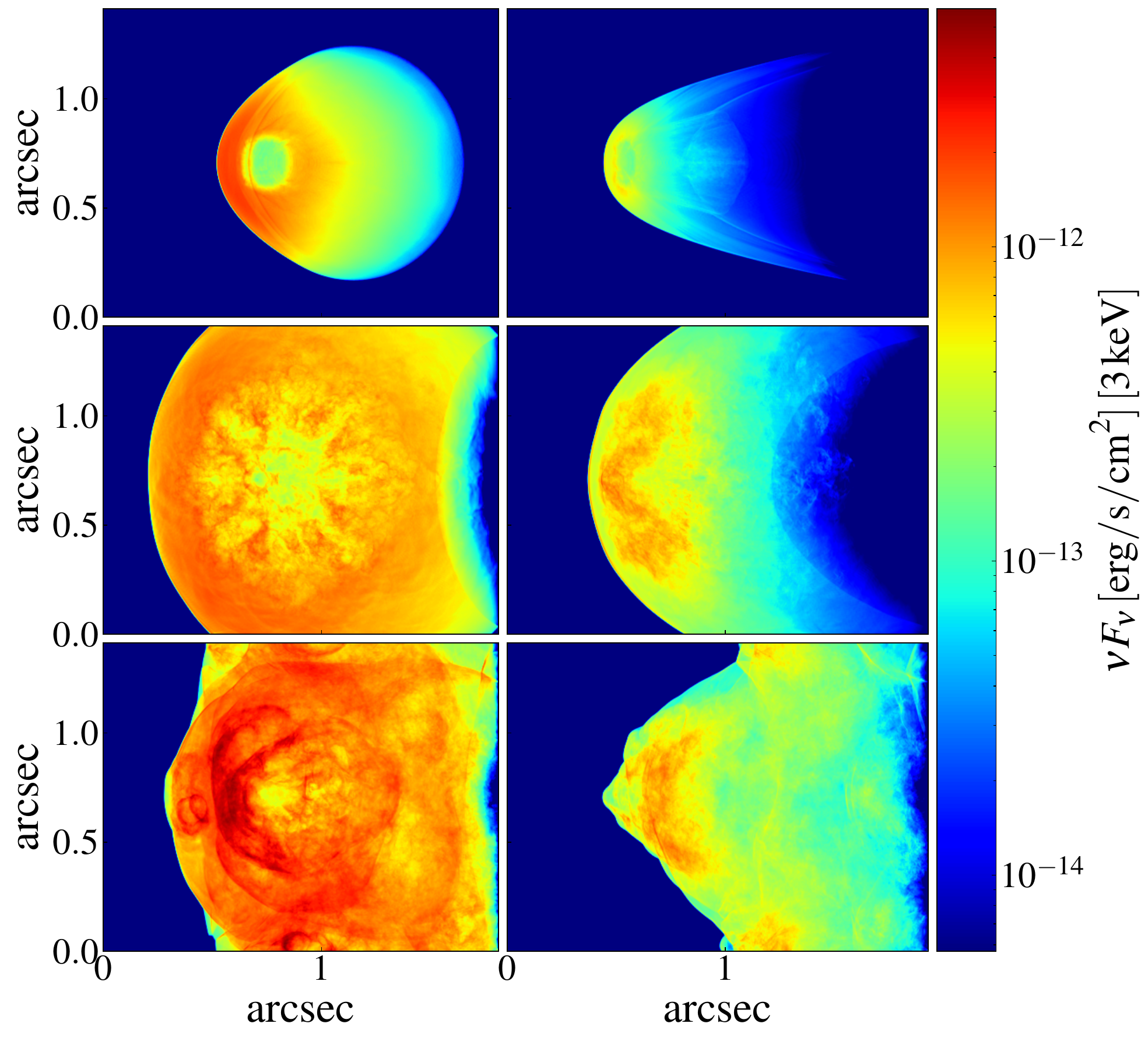}
    \caption{Same as in Fig.~\ref{2dX} but without Gaussian smoothening.}
    \label{2dX_ng}
\end{figure}

\end{document}